\documentclass[10pt]{article}
\usepackage{amsfonts}
\usepackage{pstricks}
\usepackage{pst-node}
\usepackage{epsfig}
\usepackage{epsfig}
\usepackage{graphicx}
\usepackage[font={footnotesize,it}]{caption}
\usepackage[latin5]{inputenc}
\usepackage[T1]{fontenc}
\usepackage{lipsum}
\usepackage{amsmath}
\usepackage{cite}
\begin{document}
	\def\ba{\begin{eqnarray}}
		\def\ea{\end{eqnarray}}
	\def\w{\wedge}
	\def\d{\mbox{d}}
	\def\D{\mbox{D}}
	%%%%%%%%%%%%%%%%%%%%%%%%%%%%%%%%%%%% TITLEPAGE %%%%%%%%%%%%%%%%%%%%%%%%%%%%%%
	\begin{titlepage}
		\title{Exact Solutions for Colliding Neutrino Waves and the Absence of Mutual Scattering for a Single Grassmann Mode}
		\author{Yorgo Senikoglu \footnote{ysenikoglu@gsu.edu.tr}}
		\date{%
			\small D\'{e}partement de Math\'{e}matiques, Universit\'{e} Galatasaray, 34349 Be\c{s}ikta\c{s}, \.{I}stanbul, Turkey\\
			\today}
		
		\maketitle
		
\begin{abstract}
	We study the head-on collision of two plane neutrino waves in the coupled Einstein-Weyl system, treating the massless Weyl spinor as an odd Grassmann valued field. We show that the mixed components of the Einstein equations force the two counter propagating waves to be built on a single Grassmann mode, whose bilinear $\varepsilon=\eta\eta^{*}$ is nilpotent. With this structure the full coupled system is solved exactly in the interaction region for arbitrary amplitude and phase profiles, and the metric and spinor fields are matched to the single wave regions by the non-impulsive O'Brien-Synge junction conditions. The metric functions contain no terms coupling the two waves; the Weyl scalar $\Psi_2$ vanishes, and $\Psi_0$, $\Psi_4$ are proportional to the neutrino energy fluxes. The waves therefore traverse each other without mutual scattering; we trace this to the nilpotency of the single mode bilinear. We discuss the limits of this result and indicate where cross couplings can arise.
\end{abstract}

\end{titlepage}
\newpage
%%%%%%%%%%%%%%%%%%%%%%%%%%%%%%%%%%%% INTRODUCTION %%%%%%%%%%%%%%%%%%%%%%%%%%%%%%

\section{Introduction}

The collision of plane waves is the simplest setting in which the nonlinearity of general relativity can be studied exactly. Since the seminal work of Khan and Penrose \cite{khan-penrose} and of Szekeres \cite{szekeres1,szekeres2}, and the subsequent discovery of exact closed-form solutions for colliding gravitational shocks by Nutku and Halil \cite{nutku-halil}, exact solutions are known for the collision of gravitational waves, of electromagnetic waves \cite{bell-szekeres} and of mixtures of the two \cite{griffiths0}. In these solutions the mutual focusing of the waves typically produces a curvature singularity or a horizon in the interaction region \cite{yurtsever,clarke-hayward}.

In classical general relativity the head-on collision of waves is governed by strong nonlinearities: whether one deals with gravitational waves, electromagnetic pulses, or coupled Einstein-Maxwell systems \cite{chandrasekhar-xanthopoulos}, the mutual focusing typically leads to caustics and curvature singularities in the interaction region. The situation studied here is different: the two neutrino waves pass through each other without mutual scattering. As explained in Sec.~4, the regularity of the solution is formal, a consequence of the nilpotency of the Grassmann source, and should not be read as a general mechanism preventing focusing singularities.

The corresponding problem for massless spin-$1/2$ fields was first considered by Griffiths \cite{griffiths-cmp,griffiths-jpa,griffiths0}, who started from the coupled Einstein-Weyl equations in the Newman-Penrose formalism. As discussed in \cite{dereli-senikoglu}, the collision of the so-called ghost neutrino waves could not be extended explicitly to the interaction region, and the problem has remained open. An extension to spacetimes with torsion \cite{griffiths-ec,dereli-tucker} did not lead to a consistent exact colliding solution either \cite{dereli-senikoglu}.

In this paper we revisit the colliding neutrino problem with the neutrino described as a classical Weyl field with odd Grassmann valued components, as is customary for classical fermionic fields \cite{dewitt} and in colliding-wave problems in supergravity \cite{rosenbaum,dereli-senikoglu}.

We find that the mixed components of the Einstein equations force the two counter-propagating neutrino waves to be built on one and the same Grassmann mode. Independent modes are excluded when region~I is flat. With this structure the complete coupled Einstein-Weyl system is solved exactly in the interaction region, for arbitrary amplitude and phase profiles, with fields matched to the single-wave regions by the non-impulsive O'Brien-Synge conditions \cite{obrien-synge}. The Weyl scalar $\Psi_2$ vanishes, while $\Psi_0$ and $\Psi_4$ are proportional to the neutrino energy fluxes. The two waves pass through each other without mutual scattering. We trace this to the nilpotency $\varepsilon^2=0$ of the single mode bilinear, and discuss where cross-couplings could arise.

The paper is organized as follows. Section~2 sets up the Einstein-Weyl system. Section~3 contains the solution, the junction conditions and the curvature. Section~4 collects the conclusions, and the Appendix lists the verification of the field equations.

%%%%%%%%%%%%%%%%%%%%%%%%%%%%%%%%%%%% SECTION 2 %%%%%%%%%%%%%%%%%%%%%%%%%%%%%%
\section{Theoretical Framework}

We use the plane-symmetric metric with a single polarization of the
gravitational field,
\begin{equation}
	g = 2e^{-M}\,du\,dv + e^{-U}\left(e^{V}dx^{2} + e^{-V}dy^{2}\right),
	\label{eq:metric}
\end{equation}
where $U$, $V$ and $M$ depend on $(u,v)$ only, with null coordinates defined as $u = \frac{z+t}{\sqrt{2}}$ and $v = \frac{z-t}{\sqrt{2}}$. A function of $u$ describes a wave moving in the $-z$ direction and a function of $v$ a wave moving in the $+z$ direction. The spacetime is divided into four
regions: region~I ($u<0,\ v<0$) is flat; region~II ($u>0,\ v<0$) and region~III
($u<0,\ v>0$) contain a single wave each; region~IV ($u>0,\ v>0$) is the
interaction region.

\begin{figure}[htb!]
	\centerline{\includegraphics[scale=1]{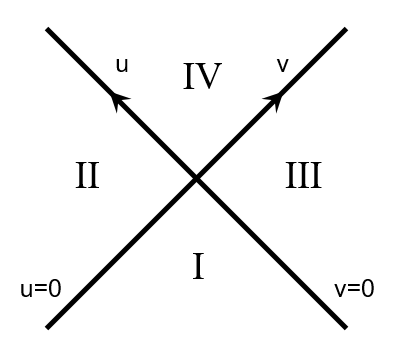}}
	\caption{Spacetime is divided into four regions as shown in the figure. Two spacelike coordinates have been suppressed.}
\end{figure}

Our analysis centers on the collision of massless, left-handed chiral neutrino field propagating in a colliding gravitational background spacetime. Because the neutrino is treated as strictly massless, its dynamics are completely captured by a two-component Weyl spinor:
\ba
\Psi =
\begin{pmatrix}
	\psi_{1} \\
	\psi_{2}
\end{pmatrix}
\ea
Within this massless framework, the propagation is governed by the two component Weyl equation without any explicit mass terms or flavor mixing effects. As a result, our findings characterize the behavior of this chiral Weyl field and do not intrinsically differentiate between Dirac and Majorana neutrino descriptions.

This spinor field satisfies the massless Weyl equation:
\ba
\sigma^{a}\nabla_{X_{a}}\Psi = 0,
\ea
where $\sigma^a :\{I ,\sigma^1, \sigma^2, \sigma^3 \}$ are the Pauli matrices, and $\nabla_{X_{a}}$ denotes the covariant derivative with respect to an orthonormal co-frame $\{e^a\}$. Consequently, the covariant exterior derivative operator is given by $\nabla = e^a \nabla_{X_{a}}$.

\noindent
The spinor components $\psi_{1}$ and $\psi_{2}$ are odd Grassmann variables and are assumed to be complex-valued functions of the spacetime coordinates $\{u,v,x,y\}$. The corresponding symmetric stress-energy tensor associated with $\Psi$ takes the form \cite{griffiths0}:
\ba \label{stress-energy}
T_{ab}[\Psi] &= \frac{i}{4}\Big(
\Psi^{\dagger}\sigma_{a}\nabla_{X_{b}}\Psi
+ \Psi^{\dagger}\sigma_{b}\nabla_{X_{a}}\Psi
- \nabla_{X_{a}}\Psi^{\dagger}\sigma_{b}\Psi
- \nabla_{X_{b}}\Psi^{\dagger}\sigma_{a}\Psi
\Big).
\ea

We introduce the complex null co-basis 1-forms
\begin{align}
l = e^{-M/2} du  , \; n = e^{-M/2} dv, \; \nonumber \\ 
m =\frac{e^{-U/2}}{\sqrt{2}}(e^{V/2} dx + e^{-V/2} i dy);
\end{align}
\noindent
The Weyl equation becomes
\ba
\psi_{1v}-\psi_1\Big(\frac{M_v}{4}+\frac{U_v}{2}\Big)&=&0, \nonumber \\
-\psi_{2u}+\psi_2\Big(\frac{M_u}{4}+\frac{U_u}{2}\Big)&=&0,
\ea
and the solution given by
\ba
\psi_1(u,v)&=&f(u)\; e^{\frac{U}{2}+\frac{M}{4}}, \nonumber \\
\psi_2(u,v)&=&h(v)\; e^{\frac{U}{2}+\frac{M}{4}}.
\ea

Solving the Weyl equation on the background \eqref{eq:metric}, the two chirality
components are described by the functions $f(u)$ and $h(v)$, which propagate along
the null directions $u=\text{const}$ (towards $-z$) and $v=\text{const}$ (towards $+z$)
respectively. They are
odd Grassmann-valued; their complex conjugates are $f^{*}$ and $h^{*}$, with the
convention $(AB)^{*}=B^{*}A^{*}$. The metric functions $U$, $V$, $M$ are real and
Grassmann-even.

The Einstein-Weyl equations then reduce to
\begin{align}
	U_{uv} - U_{u}U_{v} &= 0, \label{eq:H}\\
	2V_{uv} - U_{u}V_{v} - U_{v}V_{u} &= 0, \label{eq:G}\\
	2M_{uv} + U_{u}U_{v} - V_{u}V_{v} &= 0, \label{eq:E}\\
	ff^{*}\,V_{v} + hh^{*}\,V_{u} &= 0, \label{eq:F}\\
	i\left(-U_{uu} - M_{u}U_{u} + \tfrac12 U_{u}^{2} + \tfrac12 V_{u}^{2}\right)
	&= \frac{e^{U}}{\sqrt2}\left(f_{u}f^{*} - f f_{u}^{*}\right), \label{eq:A}\\
	i\left(U_{vv} + U_{v}M_{v} - \tfrac12 U_{v}^{2} - \tfrac12 V_{v}^{2}\right)
	&= \frac{e^{U}}{\sqrt2}\left(h_{v}h^{*} - h h_{v}^{*}\right), \label{eq:C}\\
	\tfrac12 f_{u}h^{*} + \tfrac14 f h^{*}\left(M_{u}-U_{u}\right)
	- \tfrac14 h f^{*}V_{u} &= 0, \label{eq:B}\\
	-\tfrac12 h_{v}f^{*} + \tfrac14 h f^{*}\left(U_{v}-M_{v}\right)
	+ \tfrac14 f h^{*}V_{v} &= 0. \label{eq:D}
\end{align}
Equations \eqref{eq:A} and \eqref{eq:C} are the $uu$ and $vv$ components of the
Einstein equations, with $R_{uu}=U_{uu}+M_{u}U_{u}-\tfrac12U_{u}^{2}-\tfrac12V_{u}^{2}$
and the analogous expression for $R_{vv}$; the right-hand sides are the neutrino
energy fluxes along $u$ and $v$. Equations \eqref{eq:B} and \eqref{eq:D} are the
mixed components. Because the metric \eqref{eq:metric} has $R_{ui}=R_{vi}=0$
($i=x,y$), the corresponding components of the neutrino stress tensor must vanish.
These are the equations that couple the two waves. The trace of the stress tensor
vanishes for a massless Weyl field, so $R=0$; given \eqref{eq:H}, this is
equivalent to \eqref{eq:E}.

The system is consistent in the following sense. Writing
$\mathcal E_u\equiv U_{uu}+M_{u}U_{u}-\tfrac12U_{u}^{2}-\tfrac12V_{u}^{2}$ and
$\mathcal E_v\equiv U_{vv}+U_{v}M_{v}-\tfrac12U_{v}^{2}-\tfrac12V_{v}^{2}$, the
equations \eqref{eq:H}, \eqref{eq:G} and \eqref{eq:E} imply
$\partial_v\mathcal E_u=U_v\,\mathcal E_u$ and $\partial_u\mathcal E_v=U_u\,\mathcal E_v$.
This is compatible with right-hand sides of the form $e^{U}$ times a function of $u$
only in \eqref{eq:A}, and $e^{U}$ times a function of $v$ only in \eqref{eq:C}, as
required for waves that depend on a single null coordinate.

%%%%%%%%%%%%%%%%%%%%%%%%%%%%%%%%%%%% SECTION 3 %%%%%%%%%%%%%%%%%%%%%%%%%%%%%%
\section{Solution}

\subsection{The single-mode condition}

We assume that region~I is flat and that the fields are continuous across the null
hypersurfaces $u=0$ and $v=0$. The functions $U$, $V$, $M$ then have no
Grassmann-degree-zero part: they are nilpotent in regions~II-IV.

Consider the mixed equations \eqref{eq:B} and \eqref{eq:D}. For independent odd
generators the products $fh^{*}$ and $hf^{*}$ do not vanish. The terms containing
$M_u-U_u$ and $V_u$ in \eqref{eq:B} multiply $fh^{*}$ and $hf^{*}$ by a nilpotent
field, so they are of Grassmann degree at least four, whereas $f_uh^{*}$ has a
degree-two part. The degree-two part of \eqref{eq:B} therefore gives
\begin{equation}
	f_{u}h^{*}=0,
\end{equation}
and likewise $h_vf^{*}=0$ from \eqref{eq:D}. For independent odd generators this
is possible with $f_u\neq0$ and $h\neq0$ only if $f_u$ and $h^{*}$ point along the
same odd direction: the two waves must be built on the same neutrino mode.
Independent modes are thus excluded in this setting.

This restriction is not an ansatz but a consistency condition: with independent odd
generators for the two waves, the degree-two part of \eqref{eq:B} would not vanish.
In the class of configurations considered here (flat region~I, fields continuous
across the fronts) the field equations therefore select a single Grassmann mode.

We therefore take
\begin{equation}
	f(u) = A(u)\,\eta\, e^{i\alpha(u)}, \qquad h(v) = H(v)\,\eta^{*}\, e^{i\beta(v)},
	\label{eq:neutrino}
\end{equation}
where $\eta$ is a constant odd Grassmann generator, $A(u)$ and $H(v)$ are real
amplitude profiles, and $\alpha(u)$, $\beta(v)$ are real phases. The case
$A=H=1$ is a pair of constant-amplitude waves that switch on at the fronts. Defining
\begin{equation}
	\varepsilon \equiv \eta\,\eta^{*}, \qquad \varepsilon^{2}=0,
\end{equation}
the only nonvanishing bilinears are
\begin{align}
	ff^{*} &= A^{2}\varepsilon, & hh^{*} &= -H^{2}\varepsilon, \nonumber\\
	f_{u}f^{*}-ff_{u}^{*} &= 2i A^{2}\alpha'\,\varepsilon, &
	h_{v}h^{*}-hh_{v}^{*} &= -2i H^{2}\beta'\,\varepsilon, \label{eq:bilinears}
\end{align}
while $fh^{*}=hf^{*}=f_{u}h^{*}=h_{v}f^{*}=0$ because $\eta^{2}=\eta^{*2}=0$. The
terms containing $A'$ and $H'$ cancel identically in the energy fluxes, and
$hh^{*}=-H^{2}\varepsilon$ because $\eta^{*}\eta=-\eta\eta^{*}$.
Hence \eqref{eq:B} and \eqref{eq:D} are satisfied identically for arbitrary
$U$, $V$, $M$. The neutrinos enter the remaining equations only through
$\varepsilon$, and since $\varepsilon^{2}=0$ they act linearly: \eqref{eq:F} becomes
$\varepsilon\,(A^{2}V_{v}-H^{2}V_{u})=0$, and \eqref{eq:A}, \eqref{eq:C} become
\begin{align}
	-U_{uu} - M_{u}U_{u} + \tfrac12 U_{u}^{2} + \tfrac12 V_{u}^{2}
	&= \sqrt2\,A^{2}\alpha'\,\varepsilon\, e^{U}, \label{eq:A2}\\
	U_{vv} + U_{v}M_{v} - \tfrac12 U_{v}^{2} - \tfrac12 V_{v}^{2}
	&= -\sqrt2\,H^{2}\beta'\,\varepsilon\, e^{U}. \label{eq:C2}
\end{align}
More generally, $f=a(u)W$ and $h=b(v)W^{*}$, with $W$ any odd element and $a$, $b$
even-valued functions that may carry nilpotent parts, also satisfy
\eqref{eq:B} and \eqref{eq:D} identically because $W^{2}=0$. The analysis below
carries over with $\varepsilon\to WW^{*}$ and $A^{2}\alpha'$, $H^{2}\beta'$ replaced by the
corresponding even-valued fluxes.

\subsection{Solution of the field equations}

We look for a solution of the form
\begin{equation}
	U = -V = -\ln\bigl(1+r(u)+s(v)\bigr), \qquad M = f_{2}(u)+f_{1}(v).
	\label{eq:solution}
\end{equation}
Write $\Phi\equiv 1+r+s$, so that $U_{u}=-r'/\Phi$ and $U_{v}=-s'/\Phi$.

First, equation \eqref{eq:H} holds identically for any $r(u)$ and $s(v)$, since
$e^{-U}=\Phi$ is a sum of a function of $u$ and a function of $v$.
Second, with $V=-U$, equation \eqref{eq:G} reduces to $-2(U_{uv}-U_{u}U_{v})=0$, which
is \eqref{eq:H}, and \eqref{eq:E} reduces to $2M_{uv}=0$, so
$M=f_{2}(u)+f_{1}(v)$.
Third, in \eqref{eq:A2} the shear term doubles the $U_{u}^{2}$ term, and
\begin{equation}
	-U_{uu}+U_{u}^{2}-M_{u}U_{u} = \frac{r''+f_{2}'\,r'}{\Phi}
	\quad\Longrightarrow\quad
	r'' + f_{2}'\,r' = \sqrt2\,\varepsilon\,A^{2}\alpha' ,
	\label{eq:rode}
\end{equation}
using $e^{U}=1/\Phi$. Likewise \eqref{eq:C2} gives
\begin{equation}
	s'' + f_{1}'\,s' = \sqrt2\,\varepsilon\,H^{2}\beta' .
	\label{eq:sode}
\end{equation}
Finally, equation \eqref{eq:F} reads $\varepsilon\,(A^{2}s'-H^{2}r')/\Phi=0$; it is satisfied
because $r$ and $s$ are proportional to $\varepsilon$ (see below).

Thus the whole problem reduces to the two ordinary differential equations
\eqref{eq:rode} and \eqref{eq:sode} for $r(u)$ and $s(v)$, given the neutrino
amplitudes and phases and the gauge functions $f_{1}$, $f_{2}$.

We need to put a remark on the choice $V=-U$. The relation $V=-U$ is an ansatz, not a
consequence of the field equations. Since $V_u^2$ and $V_v^2$ are $O(\varepsilon^2)$
whenever $V\propto\varepsilon$, the general solution at this order has
$V=\varepsilon[\rho(u)+\sigma(v)]$ with arbitrary gravitational-wave profiles
$\rho$, $\sigma$, while $U$ and $M$ are unchanged: the functions $r$ and $s$ obey the
same equations \eqref{eq:rode} and \eqref{eq:sode}. The choice $V=0$ describes a
pure neutrino field, and $V=-U$ describes a gravitational wave locked to the
neutrino flux.

\subsection{Junction conditions and the solution in the four regions}

We require $U$, $V$, $M$ and their first derivatives to be continuous across
$u=0$ and $v=0$, so that the fronts carry no impulsive component, and region~I to
be flat. This gives
\begin{equation}
	r(0)=r'(0)=0, \qquad s(0)=s'(0)=0, \qquad f_{2}(0)=f_{1}(0)=0 .
	\label{eq:junction}
\end{equation}
No condition on $A(0)$ or $H(0)$ is needed. Equation \eqref{eq:rode} can be written
$(e^{f_{2}}r')'=\sqrt2\,\varepsilon\, e^{f_{2}}A^{2}\alpha'$, and with
\eqref{eq:junction} it integrates to
\begin{align}
	r(u) &= \sqrt2\,\varepsilon\int_{0}^{u}\!dx\; e^{-f_{2}(x)}\int_{0}^{x}\!dy\;
	e^{f_{2}(y)}\,A(y)^{2}\alpha'(y), \label{eq:rsol}\\
	s(v) &= \sqrt2\,\varepsilon\int_{0}^{v}\!dx\; e^{-f_{1}(x)}\int_{0}^{x}\!dy\;
	e^{f_{1}(y)}\,H(y)^{2}\beta'(y). \label{eq:ssol}
\end{align}
Both are proportional to $\varepsilon=\eta\eta^{*}$, so
$\ln(1+r+s)=r+s$ and $U=-(r+s)$, $V=r+s$. In the gauge $f_{1}=f_{2}=0$,
\begin{equation}
	r(u)=\sqrt2\,\varepsilon\int_{0}^{u}(u-y)\,A(y)^{2}\alpha'(y)\,dy,
	\qquad
	s(v)=\sqrt2\,\varepsilon\int_{0}^{v}(v-y)\,H(y)^{2}\beta'(y)\,dy .
\end{equation}
The solution in each region is:
\begin{itemize}
	\item Region I ($u<0,\ v<0$): $U=V=M=0$ and $f=h=0$.
	\item Region II ($u>0,\ v<0$): $f=A\,\eta\,e^{i\alpha(u)}$, $h=0$, and
	$U_{II}=-\ln\bigl(1+r(u)\bigr)$, $V_{II}=\ln\bigl(1+r(u)\bigr)$, $M_{II}=f_{2}(u)$.
	\item Region III ($u<0,\ v>0$): $f=0$, $h=H\,\eta^{*}e^{i\beta(v)}$, and
	$U_{III}=-\ln\bigl(1+s(v)\bigr)$, $V_{III}=\ln\bigl(1+s(v)\bigr)$, $M_{III}=f_{1}(v)$.
	\item Region IV ($u>0,\ v>0$): $f=A\,\eta\,e^{i\alpha(u)}$, $h=H\,\eta^{*}e^{i\beta(v)}$, and
	$U_{IV}=-\ln\bigl(1+r(u)+s(v)\bigr)$, $V_{IV}=-U_{IV}$, $M_{IV}=f_{2}(u)+f_{1}(v)$.
\end{itemize}
With the Heaviside function $\Theta$ this is the single expression
\begin{align}
	f &= \Theta(u)\,A(u)\,\eta\, e^{i\alpha(u)}, \qquad h = \Theta(v)\,H(v)\,\eta^{*} e^{i\beta(v)},\\
	U &= -\ln\bigl[1+\Theta(u)\,r(u)+\Theta(v)\,s(v)\bigr], \qquad V=-U,\\
	M &= \Theta(u)\,f_{2}(u)+\Theta(v)\,f_{1}(v).
\end{align}
Derivatives of $\Theta$ do not contribute, since $r(0)=r'(0)=f_{2}(0)=0$ (and
likewise in $v$). A step in the neutrino amplitude at the front, $A(0)\neq0$,
produces no $\delta$-function in the energy flux, because the terms containing $A'$
cancel in \eqref{eq:bilinears}. If the neutrinos are absent ($\varepsilon=0$ or
$A=H=0$), then $r=s=0$ and all metric functions vanish, so that the solution reduces
to Minkowski spacetime in all four regions.

As an example, take constant amplitudes $A=H=1$ and monochromatic phases
$\alpha=\omega u$, $\beta=\omega' v$. In the gauge $f_1=f_2=0$ one finds
$r=\varepsilon\,\omega u^{2}/\sqrt2$ and $s=\varepsilon\,\omega' v^{2}/\sqrt2$, so that in region~IV
\begin{equation}
	U=-V=-\frac{\varepsilon}{\sqrt2}\left(\omega u^{2}+\omega' v^{2}\right),
\end{equation}
and the nonvanishing Weyl scalars of Sec.~3.4 are the constants
$\Psi_0=-\omega\varepsilon/\sqrt2$ and $\Psi_4=-\omega'\varepsilon/\sqrt2$.

\subsection{Curvature}

We use the frame vectors dual to the co-basis $(l,n,m)$, namely
$X_l=e^{M/2}\partial_u$, $X_n=e^{M/2}\partial_v$ and $X_m$, $X_{\bar m}$ along the
transverse directions, and define $\Psi_0=C(X_l,X_m,X_l,X_m)$,
$\Psi_2=C(X_l,X_m,X_{\bar m},X_n)$ and $\Psi_4=C(X_n,X_{\bar m},X_n,X_{\bar m})$,
where $C$ is the Weyl tensor. The Weyl scalars of the metric \eqref{eq:metric} are,
up to sign conventions,
\begin{align}
	\Psi_0 &= \tfrac12 e^{M}\bigl(U_uV_u-M_uV_u-V_{uu}\bigr), \nonumber\\
	\Psi_2 &= \tfrac16 e^{M}\bigl(V_uV_v+M_{uv}-U_{uv}\bigr)
	= \tfrac14 e^{M}\bigl(V_uV_v-U_uU_v\bigr), \nonumber\\
	\Psi_4 &= \tfrac12 e^{M}\bigl(U_vV_v-M_vV_v-V_{vv}\bigr),
	\label{eq:weyl}
\end{align}
and $\Psi_1=\Psi_3=0$; the second form of $\Psi_2$ follows from \eqref{eq:H}
and \eqref{eq:E}. The Ricci tensor has the components
\begin{equation}
	R_{uu}=-\sqrt2\,A^{2}\alpha'\,\varepsilon\, e^{U}, \qquad
	R_{vv}=-\sqrt2\,H^{2}\beta'\,\varepsilon\, e^{U},
\end{equation}
by \eqref{eq:A2} and \eqref{eq:C2}, while all other components vanish by
\eqref{eq:H}, \eqref{eq:G} and \eqref{eq:E}, and $R=0$. The stress tensor is
therefore that of two counter-propagating null fluxes.

For the solution \eqref{eq:solution} we have $V_uV_v=U_uU_v$, hence
\begin{equation}
	\Psi_2=0 ,\qquad
	\Psi_0=-\frac{1}{\sqrt2}\,A^{2}\alpha'\,\varepsilon\,e^{M+U},\qquad
	\Psi_4=-\frac{1}{\sqrt2}\,H^{2}\beta'\,\varepsilon\,e^{M+U}.
	\label{eq:weylsol}
\end{equation}
Since $\varepsilon\,e^{U}=\varepsilon$ and $M=f_2(u)+f_1(v)$, in the gauge
$f_1=f_2=0$ the scalar $\Psi_0$ depends on $u$ only and $\Psi_4$ on $v$ only.
In this tetrad, $\Psi_0=\tfrac12e^{M}R_{uu}$ and $\Psi_4=\tfrac12e^{M}R_{vv}$: for
$V=-U$ the Weyl curvature is locked to the neutrino energy fluxes. For the pure
neutrino field $V=0$, $\Psi_0=\Psi_4=0$ at order $\varepsilon$, and for a general
$V=\varepsilon[\rho+\sigma]$ one finds
$\Psi_0=-\tfrac12 e^{M}\varepsilon(\rho''+f_2'\rho')$ and
$\Psi_4=-\tfrac12 e^{M}\varepsilon(\sigma''+f_1'\sigma')$ with $\Psi_2=0$.

The vanishing of $\Psi_2$ does not follow from $M_{uv}=0$ alone. On shell
$\Psi_2\propto V_uV_v-U_uU_v$, which is $O(\varepsilon^{2})=0$ for any $V\propto\varepsilon$
and is zero exactly for $V=-U$. It is a necessary diagnostic of the absence of
nonlinear interaction, in contrast to the gravitino collision of
\cite{dereli-senikoglu}, where $\Psi_2\propto|\eta_1|^2|\eta_2|^2\,\theta(u)\theta(v)$
is nonzero in the interaction region.

\subsection{Discussion}

Since $\varepsilon^{2}=0$, the solution is exact in the Grassmann algebra, but it
coincides with first-order perturbation theory in $\varepsilon$ about Minkowski
spacetime. The nontrivial content is that the full nonlinear Einstein-Weyl system,
including the mixed equations \eqref{eq:B} and \eqref{eq:D}, is consistent, and that
it forces the single-mode structure which in turn removes all coupling between the
two waves.

The neutrino energy flux is proportional to $A^{2}\alpha'$ (respectively $H^{2}\beta'$),
that is to the phase gradient of the wave function weighted by the squared amplitude;
a wave of constant phase carries no energy. The second derivatives $U_{uu}$ and
$U_{vv}$ are discontinuous across the fronts, which is the energy flux switching on
with a finite value, while $U$, $V$, $M$ and their first derivatives are continuous.
Since $\varepsilon$ is nilpotent, the energy-momentum tensor components are Grassmann-even
nilpotent quantities. Consequently, statements about the sign or positivity of the energy
flux are not meaningful in this formal setting.

The functions $f_{1}$ and $f_{2}$ reflect the freedom to reparametrize the null
coordinates, $u\to\tilde u(u)$, $v\to\tilde v(v)$. Choosing them so that $r$ and
$s$ are linear in $u$ and $v$ puts the whole neutrino profile into $M$, but then
$r'(0)\neq0$ and $s'(0)\neq0$, i.e.\ $U_{u}$ and $U_{v}$ jump at the fronts. This
corresponds to an impulsive component and is excluded by \eqref{eq:junction}.

Because $\varepsilon^{2}=0$, the interaction region is a linear superposition of
the two single-wave solutions: $U$, $V$ and $M$ contain no cross term
$\Theta(u)\Theta(v)$ beyond the logarithm, which truncates after the linear term.
Equivalently, in the gauge $f_1=f_2=0$ the curvature scalars \eqref{eq:weylsol}
depend on $u$ and $v$ separately. In this sense the two neutrino waves pass
through each other without mutual scattering. This conclusion rests on two
facts: the mixed Einstein equations force a single Grassmann mode, and the
bilinear of a single mode is nilpotent. It does not depend on the choice of $V$,
and it persists if $f=a(u)W$, $h=b(v)W^{*}$ carry additional nilpotent factors,
because every neutrino-induced term still carries the same $WW^{*}$.

Cross terms can arise only if the gravitational data contain an independent even
Grassmann structure, for example a second bilinear $\varepsilon_g=\theta\theta^{*}$ in
$V$. In that case \eqref{eq:F} ties the corresponding profile of $V$ to the
neutrino fluxes, $V|_{\varepsilon_g}=k\,[\mathcal A(u)+\mathcal B(v)]$ with
$\mathcal A=\int_0^uA^{2}$ and $\mathcal B=\int_0^vH^{2}$, and at order
$\varepsilon\varepsilon_g$ the equations \eqref{eq:G} and \eqref{eq:E} acquire
products of $u$- and $v$-dependent terms. For waves whose amplitude does not vanish
at the fronts this would require a jump of $V_u$ there, since
$\mathcal A'(0)=A(0)^{2}\neq0$, i.e.\ an impulsive gravitational component. For
amplitude profiles with $A(0)=H(0)=0$ the junction conditions can be satisfied. We do
not pursue this case here.

\section{Conclusion}
We have found an exact solution of the coupled Einstein-Weyl equations describing the head-on collision of two plane neutrino waves with arbitrary amplitude and phase profiles, with the Weyl spinor treated as a Grassmann-valued field and the fields matched across the null hypersurfaces by the non-impulsive O'Brien-Synge conditions. The mixed Einstein equations require the two waves to be built on a single Grassmann mode, and the nilpotency of its bilinear, $\varepsilon^2=0$, then removes every term that would couple the two waves. Consequently, with the backreaction of the neutrino energy-momentum tensor included to all orders allowed by the algebra, the interaction region is an exact linear superposition of the single-wave solutions. We stress that this is a consequence of the single Grassmann mode: independent source modes are excluded by the field equations when region~I is flat, and cross-couplings would have to come from additional even Grassmann structure in the gravitational sector.

The solution is regular only in a formal sense. The transverse factor $1+r(u)+s(v)$ cannot vanish, and all curvature components are finite, because they are linear in the nilpotent $\varepsilon$ and the body of the metric is flat. This should not be read as a physical mechanism preventing the focusing singularities of bosonic colliding-wave spacetimes: the same geometry with commuting functions $r$, $s$ has $\Phi\to0$ with divergent curvature.

The curvature is characterized by $\Psi_2=0$ and by $\Psi_0$, $\Psi_4$ proportional to the neutrino energy fluxes. The vanishing of $\Psi_2$ reflects the absence of the nonlinear interaction wake, and contrasts with the gravitino collision \cite{dereli-senikoglu}, where $\Psi_2$ is nonzero in the interaction region. The relation $V=-U$ locks the gravitational-wave content to the neutrino flux; other choices of $V$ give the same $U$ and $M$ and different $\Psi_0$, $\Psi_4$.

It would be interesting to extend the analysis to curved backgrounds, to additional even Grassmann structure in the gravitational sector, and to massive neutrinos, where the stress tensor is no longer traceless.

\section*{Data Availability Statement}
No new data were created or analyzed in this study.
\bigskip

\section*{Funding Statement}
The author received no financial support for the research, authorship or publication of this article.

\end{document}